\documentclass[letterpaper]{article} 
\usepackage{aaai23}  
\usepackage{times}  
\usepackage{helvet}  
\usepackage{courier}  
\usepackage[hyphens]{url}  
\usepackage{graphicx} 
\usepackage{natbib}  
\usepackage{caption} 
\usepackage{algorithm}
\usepackage{algorithmic}
\usepackage{amsfonts}
\usepackage{amsmath}
\usepackage{xcolor} 
\usepackage{multirow}
\usepackage{booktabs}

\usepackage{newfloat}
\usepackage{listings}
\DeclareCaptionStyle{ruled}{labelfont=normalfont,labelsep=colon,strut=off} 
\floatstyle{ruled}
\newfloat{listing}{tb}{lst}{}
\floatname{listing}{Listing}
\title{Steganography of Steganographic Networks}
\author {
    Guobiao Li,
    Sheng Li\thanks{Corresponding authors},
    Meiling Li, 
    Xinpeng Zhang\footnotemark[1], 
    Zhenxing Qian 
}
\affiliations {
    Fudan University \\
    \{gbli20, lisheng, mlli20, zhangxinpeng, zxqian\}@fudan.edu.cn
}

\usepackage{bibentry}

\begin{document}

\maketitle

\begin{abstract}
Steganography is a technique for covert communication between two parties. With the rapid development of deep neural networks (DNN), more and more steganographic networks are proposed recently, which are shown to be promising to achieve good performance. Unlike the traditional handcrafted steganographic tools, a steganographic network is relatively large in size. It raises concerns on how to covertly transmit the steganographic network in public channels, which is a crucial stage in the pipeline of steganography in real world applications. To address such an issue, we propose a novel scheme for steganography of steganographic networks in this paper. Unlike the existing steganographic schemes which focus on the subtle modification of the cover data to accommodate the secrets. We propose to disguise a steganographic network (termed as the secret DNN model) into a stego DNN model which performs an ordinary machine learning task (termed as the stego task). During the model disguising, we select and tune a subset of filters in the secret DNN model to preserve its function on the secret task, where the remaining filters are reactivated according to a partial optimization strategy to disguise the whole secret DNN model into a stego DNN model. The secret DNN model can be recovered from the stego DNN model when needed. Various experiments have been conducted to demonstrate the advantage of our proposed method for covert communication of steganographic networks as well as general DNN models. 
\end{abstract}

\begin{figure}[t]
	\centering
	\includegraphics[width=1.00\linewidth]{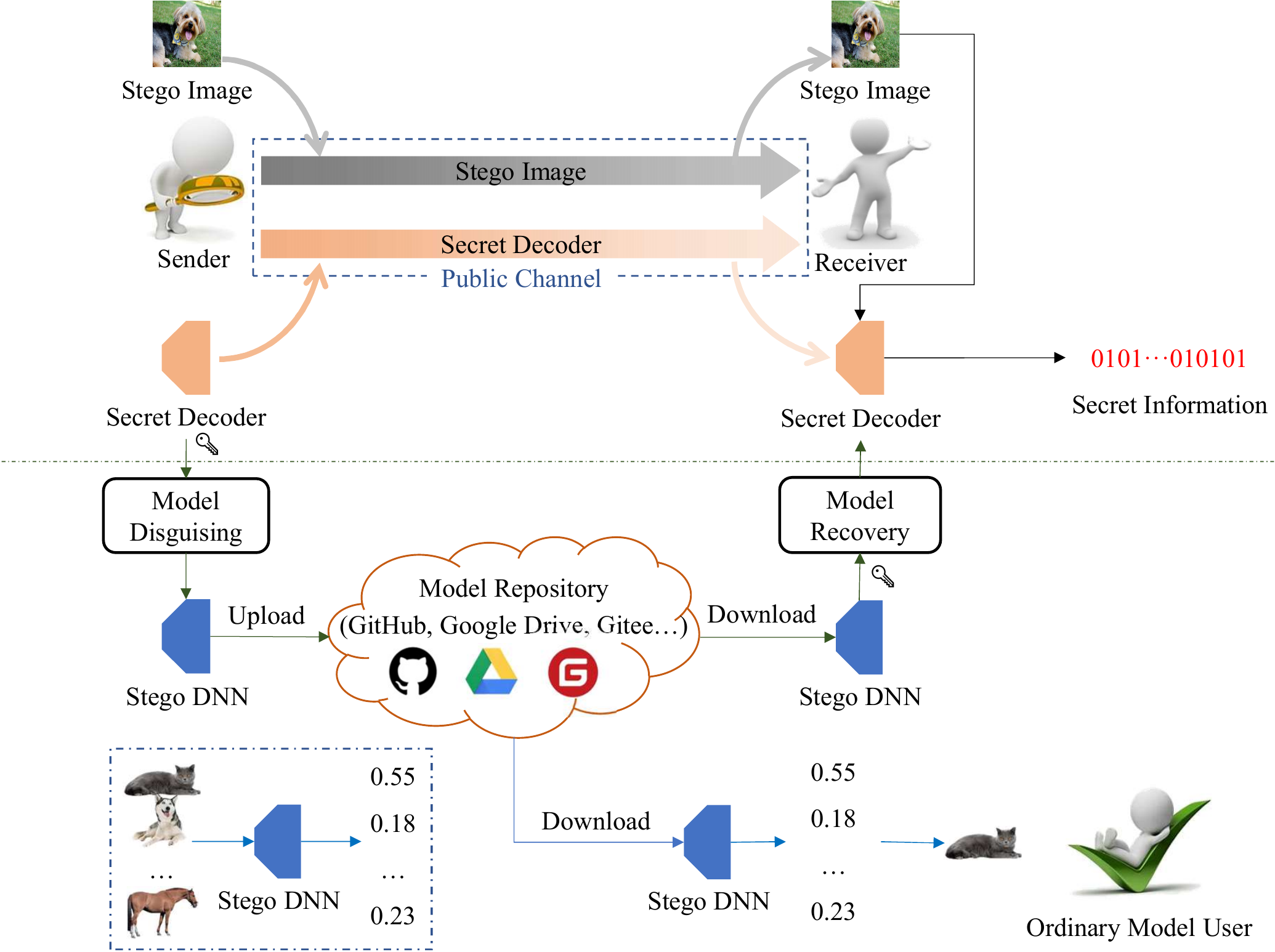}	\caption{The application scenario of the proposed method for steganography of steganographic networks.} 
	\label{fig:illustration}
\end{figure}

\section{Introduction}
Steganography is a technique which takes advantage of the redundancy in cover media to conceal secret information. It is one of the main strategies for covert communication. Various types of cover media have been considered in literature for steganography, including image \cite{8789545, 9124671}, text \cite{liu20116security}, audio \cite{8626153}, video \cite{zhang2016steganalytic} or 3D Mesh \cite{zhou2019distortion}. Traditional steganographic schemes focus on designing handcrafted algorithms to modify the cover media subtly for data embedding, which are difficult to achieve a good balance between the undetectability and embedding capacity (termed as the capacity for short). Recently, some deep neural network (DNN)-based steganographic schemes have been proposed \cite{baluja2019hiding, lu2021large, jing2021hinet}, which achieve superior performance to handcrafted ones. 

A DNN-based steganographic scheme usually takes advantage of the encoder-decoder structure of DNN for secret data embedding and extraction, which contains two steganographic networks: a DNN-based secret encoder and a DNN-based secret decoder (termed as the secret encoder and secret decoder for short). The secret encoder hides the secret into a cover media to obtain a stego media (i.e., media with hidden data), while the secret decoder recovers the secret from the stego media. Despite the advantage, we find that it is not convenient to deploy the DNN-based steganographic schemes in real world applications. Let's take image steganography as an example, when the receiver receives the stego image, he could only conduct the secret recovery with the possession of the secret decoder, as shown in the upper part of Fig. \ref{fig:illustration}. Unlike the handcrafted steganographic tools, a DNN-based secret decoder is relatively large in size. It would not be secure for the sender and receiver to conduct another transmission for the delivery of the secret decoder.

One possible solution is to treat the secret decoder as ordinary data and hide them into popular cover media such as images or videos using the existing steganographic schemes. Since the size of a secret decoder is relatively large in terms of secret information, using traditional steganographic schemes would result in the need of a large amount of cover media data due to the limited capacity, which would cause a lot of communication burden. Some DNN-based steganographic schemes \cite{baluja2019hiding, weng2019high,jing2021hinet} are able to achieve a high capacity. However, they are designed specifically for hiding images with errors introduced during the secret recovery. It is not appropriate to apply them on the hiding of secret decoder which will be of no use after the recovery.

To deal with the issue of covert communication of the secret decoder, in this paper, we propose a novel scheme for steganography of steganographic networks, where we disguise a secret DNN model (i.e., the secret decoder) into a stego DNN model which performs an ordinary machine learning task (termed as a stego task). As shown in the lower part of Fig. \ref{fig:illustration}, the sender conducts the model disguising using our proposed method to generate a stego DNN model from the secret DNN model and a key. The stego DNN model is then uploaded to public model repositories such as the github, google drive and etc. The receiver downloads the stego DNN model and performs the model recovery using the key to restore the secret decoder. Other users (those without the key) could download the stego DNN model for an ordinary machine learning task. Compared with the existing steganographic schemes, our proposed method is tailored for the covert communication of DNN models, which has the following advantages:
\begin{itemize}
  \item [1)] The communication burden hardly increases before and after steganography, and there is no need to collect a large amount of cover data to accommodate the secret DNN model.
  \item [2)] The behaviour of covert communication is well protected by placing another functional DNN model for an ordinary machine learning task. The stego DNN model could be placed in the model repository that is public available without being noticed.
\end{itemize}

To disguise the secret DNN model, we first select a set of filters according to the corresponding gradients when the model is back-propagated on the datasets for secret and stego tasks. Based on the selected filters, we establish a stego DNN model from the secret DNN model by a partial optimization strategy. The aforementioned process are progressively done to obtain a stego DNN model with good performance on the secret and stego tasks. The main contributions of this paper are summarized below. 

\begin{itemize}
    \item [1)] We tackle the problem of the covert communication of the steganographic network by disguising a secret DNN model into a stego DNN model without causing notice.
    \item [2)] We propose a partial optimization scheme for model disguising, where a stego DNN model is trained based on the secret DNN model for both the secret and stego tasks.
    \item [3)] We propose a progressive model disguising strategy for good performance of the stego DNN model.
\end{itemize}
 
\section{Related Works}
According to whether DNN has been utilized for data embedding or extraction, steganographic schemes could be mainly categorized into traditional steganography and DNN-based steganography. In this section, we give a brief review regarding these two types of steganographic schemes.

\subsection{Traditional Steganography} 
Traditional steganography usually involves the alteration of cover data using hand-craft data hiding algorithms. Liao \textit{et al.} \cite{9124671} take a batch of images as the cover data for embedding the secret message, where the image texture features are measured and considered as an indicator to determine the capacity of each image. Zhang \textit{et al.} \cite{zhang2016steganalytic} propose a video steganography scheme by exploring the motion vector for data embedding. Lu \textit{et al.} \cite{8789545} conduct data embedding on a halftoned image, where the histgram of the pixels is slightly modified to accommodate the secret message. Zhou \textit{et al.} \cite{zhou2019distortion} propose to convey secret data through 3D meshes, where the importance of different regions in the 3D meshes is measured for adaptive data embedding. To ensure the undetectability, the capacity of these schemes are limited. Taking the image steganography as an example, each pixel in a grayscale image (8 bits) could accommodate 1 bit of secret message at most for high undetectability.


\subsection{DNN-based Steganography}
Most of the DNN-based steganographic schemes are proposed for image steganography. Some works take advantage of DNN to compute the embedding cost of each pixel \cite{tang2017automatic, tang2020automatic}, which still use handcrafted algorithms for data embedding when the embedding costs are ready. 

Recently, more and more DNN-based steganographic schemes are proposed without any handcrafted processes, which usually take advantage of the encoder-decoder structure of DNN for data embedding and extraction. Zhu \textit{et al.} \cite{zhu2018hidden} pioneer the research of such techniques, where the secrets are embedded into a cover image using an end-to-end learnable DNN. Motivated by \cite{zhu2018hidden}, some researchers propose to hide secret images into cover images for high capacity data embedding \cite{baluja2019hiding, lu2021large, jing2021hinet,weng2019high}. The concept of hiding images into images is initially proposed by Baluja \textit{et al.} \cite{baluja2019hiding}, where a secret image is concatenated with the cover image into a set of different channels to learn an image encoding network for generating the stego image. The stego image is then fed into an image decoding network to extract the secret image. The image encoding and decoding networks are jointly learnt for minimized extraction errors and image distortion. Similarly, the works in \cite{jing2021hinet} and \cite{lu2021large} make use of the invertible neural network for image encoding and decoding, where the number of the channels in the secret image branch can be increased to remarkably improve the capacity. Weng \textit{et al.} \cite{weng2019high} propose to hide a secret frame into a cover frame by a temporal residual modeling technique for video steganography. Despite the advantage, the works in \cite{baluja2019hiding, lu2021large, weng2019high, jing2021hinet} are suitable for secrets in terms of images which can not be losslessly recovered in decoding.

Despite the progress in the area of steganography, little attention has been paid to the problem of covert communication of DNN models, which plays an important role in real world applications of the DNN-based steganographic schemes. In this paper, we try to tackle this problem by a network to network steganographic mechanism, which disguises the secret DNN model into a stego DNN model performing an ordinary machine learning task. The secret DNN model can be recovered from the stego DNN model with little performance degradation, while the stego DNN model is able to achieve high fidelity and undetectability.

\section{Problem Formulation}
\label{sec:3_1}
\begin{figure}[ht]
	\centering
	\includegraphics[width=0.9\linewidth]{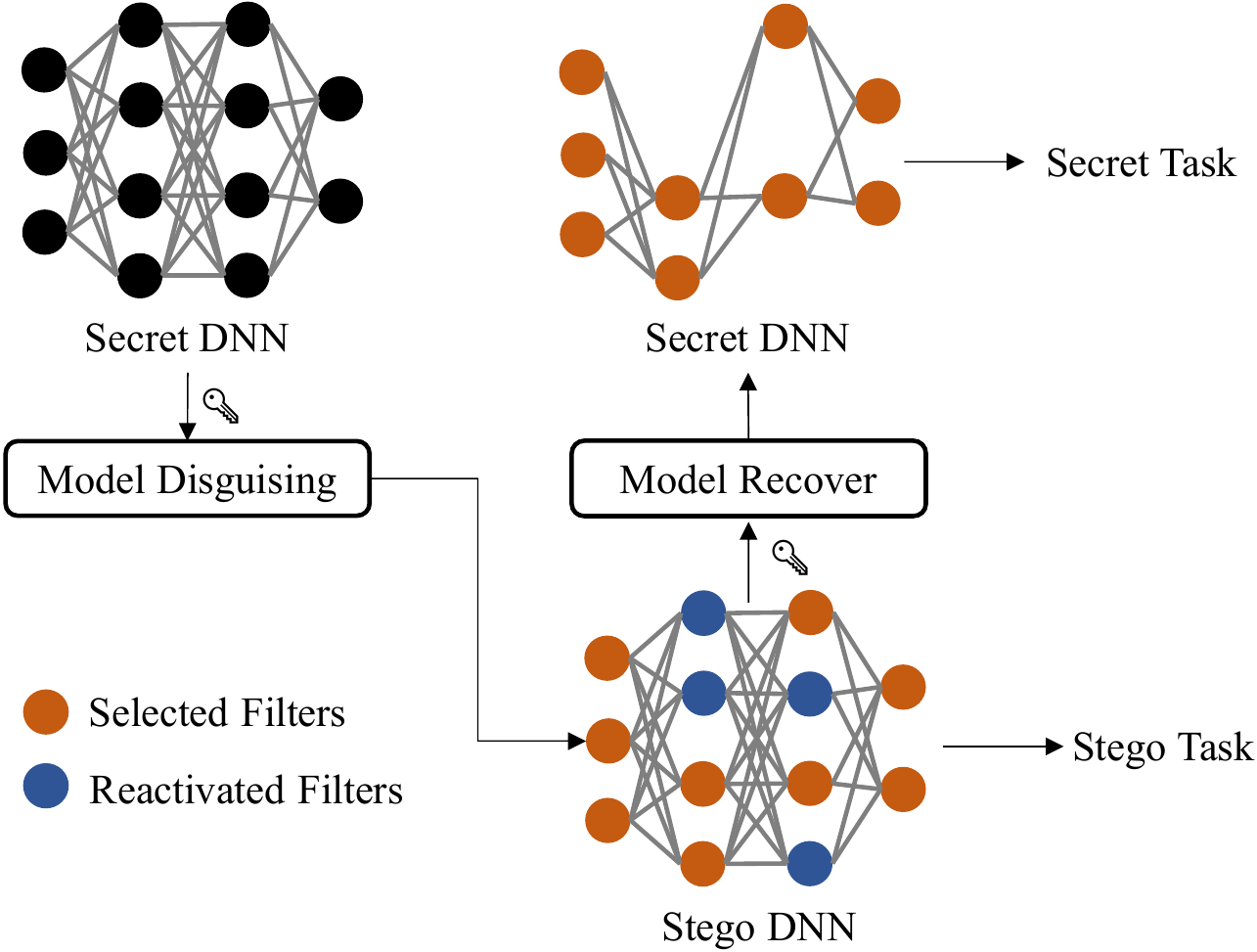}
	\caption{An illustration for the DNN model disguising and recovery.}
	\label{fig:PF}
\end{figure}

Given a secret DNN model to be disguised, our goal is to select and tune a subset of filters to preserve the function of the secret DNN, where the remaining filters are reactivated such that the whole model works on an ordinary machine learning task (i.e., a stego task) to establish a stego DNN model for model disguising. The secret DNN model could be recovered from the stego DNN model when necessary, as shown in Fig. \ref{fig:PF}.

Let's denote the secret and stego DNN models as ${\Theta}$ and $\tilde{\Theta}$, where $\tilde{\Theta}$ is an optimization of ${\Theta}$ after the model disguising. We use two datasets, a secret dataset and a stego dataset to conduct the model disguising for the secret and stego tasks, which are denoted as $D_{se}=\{\mathbf{x}_e, \mathbf{y}_e\}$ and $D_{st}=\{\mathbf{x}_t, \mathbf{y}_t\}$, with $\mathbf{x}_e$, $\mathbf{x}_t$ being the samples and $\mathbf{y}_e$, $\mathbf{y}_t$ being the labels. We further denote the model disguising process as $\mathcal{P}$ which can be formulated below
 
\begin{equation}
\begin{aligned} \label{eq1}
& \tilde{\Theta} = \mathcal{P}(\Theta, D_{se}, D_{st}, \mathcal{K}) \\
&\begin{array}{r@{\quad}r@{}l@{\quad}l}
s.t. & \left\{\begin{array}{lc}
    \Theta^\mathrm{S}\in \tilde{\Theta}  \\
    \mathcal{F}(\Theta^\mathrm{S}, \mathbf{x}_{e}) \to \mathbf{y}_{e}  \\
    \mathcal{F}(\tilde{\Theta}, \mathbf{x}_{t}) \to \mathbf{y}_{t} 
    \end{array}\right. ,
\end{array}
\end{aligned}
\end{equation}
where $\Theta^\mathrm{S}$ is the selected subset for the secret task and $\mathcal{K}$ is a key. With the stego DNN model available, the receiver can recover the secret DNN model by  
\begin{equation}
\Theta^\mathrm{S} = \mathcal{Q}(\tilde{\Theta}, \mathcal{K}),
\label{eq2}
\end{equation}
where $\mathcal{Q}$ refers to the model recovery.

The design of $\mathcal{P}$ and $\mathcal{Q}$ should satisfy the following properties for covert communication.
\begin{itemize}
  \item [1)] \textbf{Recoverability}: the performance of the recovered secret DNN model should be similar to its original version on the secret task.
  \item [2)] \textbf{Fidelity}: the performance of the stego DNN model should be high for the stego task.
  \item [3)] \textbf{Capacity}: the size of the stego DNN model should not expand too much compared with the secret DNN model for efficient transmission. 
  \item [4)] \textbf{Undetectability}: it should be difficult for the attackers to identify the existence of the secret DNN model from the stego DNN model.
\end{itemize}

\section{The Proposed Method}
\label{sec3}
Our model disguising starts with selecting a subset of filters from the convolutional layers in the secret DNN model, which is important on the secret task but trivial on the stego task. Then, we propose a partial optimization strategy to obtain a stego DNN model from the secret DNN model. The aforementioned process can be progressively performed for optimal performance. We also propose several strategies to deal with the parameters in other layers for real world applications.

\subsection{Filter Selection}
\label{sec:3_2}
Assume the secret DNN model (i.e., ${\Theta}$) contains $L$ convolutional layers, the filters in the $l$-th convolutional layer can be represented as a 4-dimensional tensor: $\mathcal{W}^{l} \in \mathbb{R}^{d^{l} \times c^{l} \times s^{l}_1 \times s^{l}_2 }$, where $d^{l}$ is the number of filters and each filter contains $c^{l}$ 2-dimensional spatial kernels. $s^{l}_1$ and $s^{l}_2$ correspond to the kernel's height and width, respectively. Denote $\mathcal{W}^{l}_{i,:,:,:}$ as the $i$-th filter and $\mathcal{W}^{l}_{:,j,:,:}$ as the $j$-th channel of all the filters in the $l$-th layer. Ignoring fully connected layer, we have: $\Theta = \{\mathcal{W}^{1}, \mathcal{W}^{2}, \cdots, \mathcal{W}^{L}\}$, where $\mathcal{W}^{l} = \{ \mathcal{W}^{l}_{1,:,:,:}, \mathcal{W}^{l}_{2,:,:,:}, \cdots, \mathcal{W}^{l}_{d^{l},:,:,:}\}$ filter-wise or $\mathcal{W}^{l} = \{ \mathcal{W}^{l}_{:,1,:,:}, \mathcal{W}^{l}_{:,2,:,:}, \cdots, \mathcal{W}^{l}_{:,c^{l},:,:}\}$ channel-wise.

The purpose of the filter selection is to choose the filters that are important on the secret task but trivial on the stego task. In our discussions, we term such filters as the important filters on model disguising. For a single machine learning task, researches have indicated that the importance of the filter could be measured according to the gradients of the filter weights \cite{8704878, yang2021robust}. In our case, however, we want to identify the filters which contain weights with large gradients on the secret task and with small gradients on the stego task. For the $i$-th filter in the $l$-th convolutional layer of a secret DNN model $\Theta$, we separately measure its importance on the secret task and stego task by $GoE^{l}_{i}$ and $GoT^{l}_{i}$, where 

\begin{equation}
\begin{aligned} \label{eq3}
&\begin{array}{r@{\quad}r@{}l@{\quad}l}
& \left\{\begin{array}{lc}
    GoE^{l}_{i} = g(\mathcal{L}_e, \mathcal{W}^{l}_{i, :, :, :}, \mathbf{x}_{e}, \mathbf{y}_{e}) \\ 
    \qquad \quad \   + \ g(\mathcal{L}_e, \mathcal{W}^{l+1}_{:, i, :, :}, \mathbf{x}_{e}, \mathbf{y}_{e})  \\
     GoT^{l}_{i} = g(\mathcal{L}_t, \mathcal{W}^{l}_{i, :, :, :}, \mathbf{x}_{t}, \mathbf{y}_{t}) \\ 
    \qquad \quad \   + \ g(\mathcal{L}_t, \mathcal{W}^{l+1}_{:, i, :, :}, \mathbf{x}_{t}, \mathbf{y}_{t}),
    \end{array}\right.
\end{array}
\end{aligned}
\end{equation}
where 
\begin{equation}
g(\mathcal{L}, \mathbf{w}, \mathbf{x}, \mathbf{y}) = \mathcal{AVG} \begin{pmatrix} \begin{vmatrix} \frac{\partial \mathcal{L}(\mathcal{F}(\Theta, \mathbf{x}), \mathbf{y})}{\partial \mathbf{w}} \end{vmatrix} \end{pmatrix},
\label{eq4}
\end{equation}



where $\mathcal{AVG}$ returns the mean of a 3-dimensional tensor, $\mathcal{L}_e$ and $\mathcal{L}_t$ are the loss functions for the secret and stego task, respectively. Here, we take the correlation between two consecutive convolutional layers into consideration. The reason is that the feature map generated by the filter $\mathcal{W}^{l}_{i,:,:,:}$ in the $l$-th layer will be convolved with the channels $\mathcal{W}^{l+1}_{:,i,:,:}$ in the ($l$+1)-th layer. There is a strong correlation between the weights in $\mathcal{W}^{l}_{i,:,:,:}$ and $\mathcal{W}^{l+1}_{:,i,:,:}$ for a certain task. Thus, we consider both the gradients of the weights in the current filter and those in the corresponding channel of the next layer. The importance of the filter on model disguising is then computed as 
\begin{equation}
\alpha_i^{l}=GoE^{l}_{i} - \lambda_{g} GoT^{l}_{i},
\label{eq5}
\end{equation}
where $\lambda_{g}$ is a weight for balance. According to $\alpha_i^{l}$, we select the top $N$ most important filters for model disguising from the secret DNN model $\Theta$, which forms the subset $\Theta^\mathrm{S}$ for performing the secret task.

\subsection{Partial Optimization}
The purpose of partial optimization is to make $\Theta^\mathrm{S}$ work on the secret task and $\Theta$ work on the stego task. First of all, we fine-tune the filters in $\Theta^{\mathrm{S}}$ on the secret dataset by 
\begin{equation}
    \Theta^{\mathrm{S}} = \Theta^{\mathrm{S}} - \lambda_{e} \nabla_{\Theta^{\mathrm{S}}} \mathcal{L}_{e}(\Theta^{\mathrm{S}}, D_{se}),
\label{eq6}
\end{equation}
where $\lambda_e$ is the learning rate of the secret task. Then, we freeze $\Theta^{\mathrm{S}}$, reinitialize and reactivate the remaining filters in $\Theta$ on the stego dataset $D_{st}$ for the following optimization problem:


\begin{equation}
 \min \limits_{\Theta} \sum \limits_{(\mathbf{x}_t, \mathbf{y}_t) \in D_{st}}  \mathcal{L}_{t}(\mathcal{F}(\Theta ,\mathbf{x}_t), \mathbf{y}_t), \quad
s.t. \  \Theta^{\mathrm{S}} \in \mathbb{C},
\label{eq7}
\end{equation}

where $\mathbb{C}$ is the constant space. To do so, we introduce $\mathrm{M}$, a binary mask with the same size as that of $\Theta$, for partial optimization, which prevents $\Theta^\mathrm{S}$ from updating during the backpropagation. Let $\lambda_t$ be the learning rate of the stego task and $\odot$ denotes the element-wise product, the secret DNN model is optimized below to form a stego DNN model $\tilde{\Theta}$:
\begin{equation}
\begin{aligned} \label{eq8}
& \Theta = \Theta - \lambda_{t} \ \mathrm{M} \odot \ \nabla_{\Theta} \mathcal{L}_{t}(\Theta, D_{st}), \\
\end{aligned}
\end{equation}
where

\begin{equation}
\begin{aligned} \label{eq9}
&\begin{array}{r@{\quad}r@{}l@{\quad}l}
& \mathrm{M}[i] = \left\{\begin{array}{lc}
    0,\quad & \Theta[i] \in \Theta^{\mathrm{S}}  \\
    1,\quad & else \end{array}\right. ,
\end{array}
\end{aligned}
\end{equation}
where $\Theta[i]$ refers to the $i$-th parameter in $\Theta$.

\begin{algorithm}[tb]
\caption{Progressive Model Disguising Algorithm.}
\label{alg:DDPA}
\textbf{Input}: Secret DNN $\Theta$,  Dataset $D_{se}$,  $D_{st}$, Thresholds $\tau_{se}$, $\tau_{st}$ \\
\textbf{Output}: Stego DNN $\tilde{\Theta}$
\begin{algorithmic}[1] 
    \STATE $t \gets 1$, $\Theta^{\mathrm{S}}_0 \gets \Theta$, $\tilde{\Theta}_0 \gets \Theta$, $\alpha_{0}^{se} \gets 0$
    \STATE \textbf{while}  $\alpha_{t-1}^{se} < \tau_{se}$ \textbf{do}
    \STATE $\qquad$ Calculate the importance of filters for filter selection from $\Theta^{\mathrm{S}}_{t-1}$ by Eq.~(\textcolor[rgb]{1,0,0}{\ref{eq5}}) 
    \STATE $\qquad$ $\Theta^{\mathrm{S}}_t \gets \mathrm{Top} \  P_{t}$ important filters in $\Theta^{\mathrm{S}}_{t-1}$
    \STATE $\qquad$ Generate mask $\mathrm{M}$ 
    \STATE $\qquad$ Reinitialize the remaining filters
    \STATE $\qquad$ Conduct the model disguising according to Eq.~(\textcolor[rgb]{1,0,0}{\ref{eq6}}) and Eq.~(\textcolor[rgb]{1,0,0}{\ref{eq8}}) to get a stego DNN model $\tilde{\Theta}_t$
    \STATE $\qquad$ \textbf{if} $\alpha_t^{st} < \tau_{st}$ \textbf{do} 
    \STATE $\qquad$ $\qquad$ Break
    \STATE $\qquad$ \textbf{end if}
    \STATE $\qquad$ $t \gets t+1$
    \STATE \textbf{end while} \\
    \STATE $\tilde{\Theta}=\tilde{\Theta}_t$
\end{algorithmic}
\end{algorithm}


\subsection{Progressive Model Disguising}
To train a disguised stego DNN model from the secret DNN model with good performance for the secret and stego tasks, we propose to progressively select the important filters $\Theta^{\mathrm{S}}$ from the secret DNN model $\Theta$ for model disguising. In particular, we iteratively select a set of important filters from  $\Theta^{\mathrm{S}}$ to train the stego DNN model, which is performed until the stego DNN model achieves satisfactory performance on the secret and stego tasks. 

For simplicity, we denote the important filter set as $\Theta_t^{\mathrm{S}}$ and the disguised stego DNN model as $\tilde{\Theta}_t$ in the $t$-th iteration, where we set $\Theta_0^{\mathrm{S}}=\Theta$ and $\tilde{\Theta}_0=\Theta$ for initialization. In the $t$-th iteration, we select the top $P_{t}$ most important filters from $\Theta_{t-1}^{\mathrm{S}}$ to form $\Theta_{t}^{\mathrm{S}}$ for training a stego DNN model $\tilde{\Theta}_{t}$, where 
\begin{equation}
  P_{t} = (\lambda_{p})^{t}\times V,
\label{eq10}
\end{equation}
where $\lambda_{p}<1$ is a decay factor and $V$ is the number of filters in $\Theta$. We terminate the iteration until the performance reduction of the stego DNN model is more than $\tau_{se}$ on the secret task, or less than $\tau_{se}$ and $\tau_{st}$ on the secret and stego tasks, respectively. Algorithm \ref{alg:DDPA} gives the pseudo code of the aforementioned process, where $\alpha_t^{se}$ and $\alpha_t^{st}$ are the performance reduction of the stego DNN model in the $t$-th iteration on the secret and stego tasks.

After the progressive model disguising, we record the property of each filter in $\tilde{\Theta}$ into $L$ binary streams $\mathbf{B}=\{b^1, b^2, \cdots, b^L\}$. Each convolutional layer corresponds to a $d^{l}$-bits binary stream $b^l$ with $d^{l}$ being the number of filters in this layer. Each bit corresponds to a filter with $0$ indicating that it belongs to $\Theta^{\mathrm{S}}$ and $1$ otherwise. We consider $\mathbf{B}$ as side information and randomly select a set of parameters from the stego DNN model according to a key $\mathcal{K}$ to host $\mathbf{B}$. In particular, we embed each bit of $\mathbf{B}$ into a selected parameter as follows. We transform the parameter into an integer and flip its least significant bit according to the bit to be hidden. 
\begin{figure}[ht]
	\centering
	\includegraphics[width=0.9\linewidth]{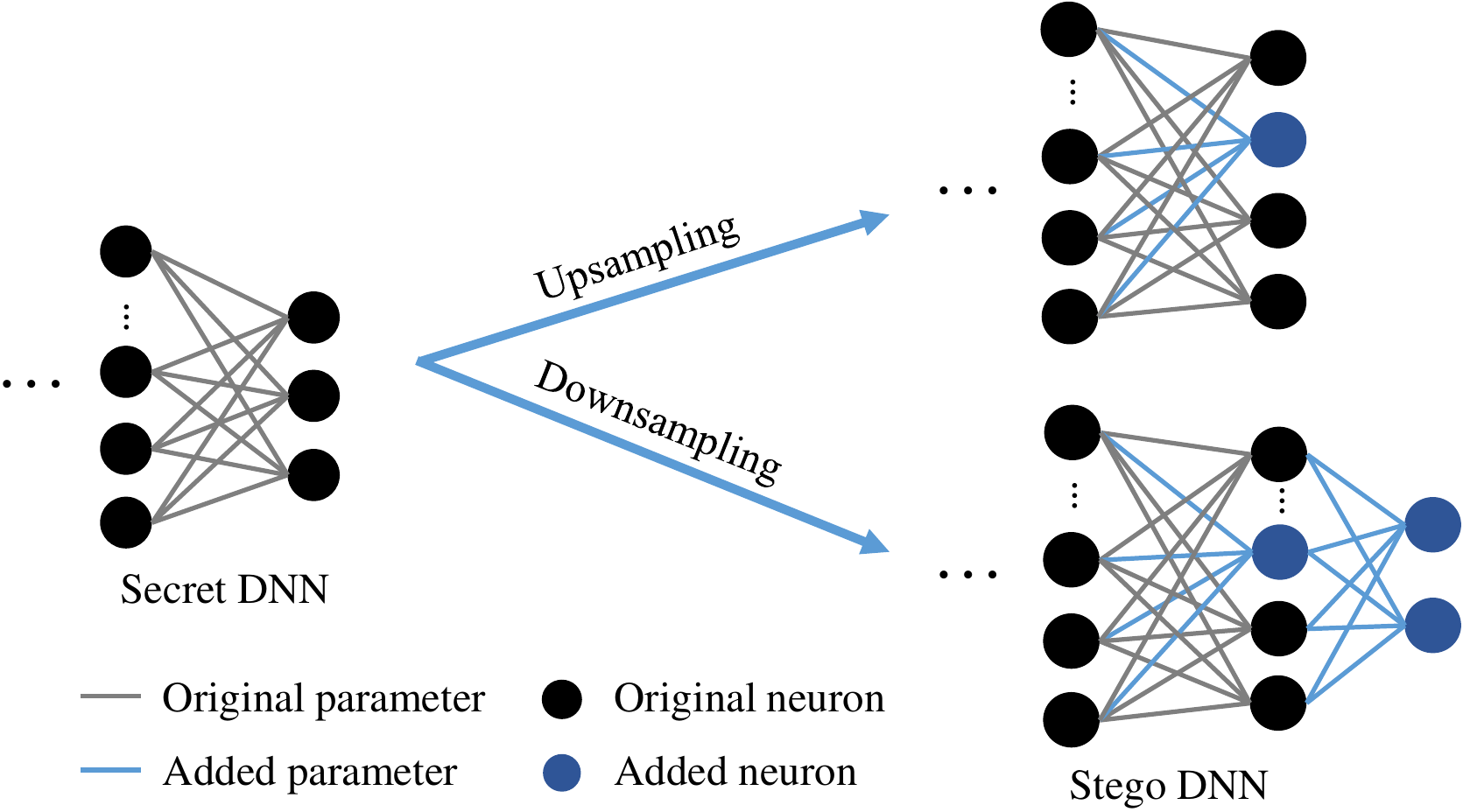}
	\caption{Output layer adaptation.}
	\label{fig:op}
\end{figure}
The parameters with hidden data are then inversely transformed into the floating numbers to complete the embedding process.

In model recovery, the receiver uses the key $\mathcal{K}$ to recover the side information $\mathbf{B}$ from the stego DNN model, based on which we can identify the important filters $\Theta^{\mathrm{S}}$ to establish a secret DNN model for the secret task.

\subsection{Hiding other layers}

\textbf{Batch Normalization Layer}.~Normalization layers, such as Batch Normalization (BN) \cite{ioffe2015batch}, Group Normalization (GN) \cite{wu2018group}, Layer Normalization (LN) \cite{ba2016layer} and Instance Normalization (IN) \cite{ulyanov2016instance}, play a crucial role in accelerating the convergence of model training and performance boosting. All these normalization layers will normalize the feature maps of the intermediate layers according to some feature statistics. The BN utilizes the moving averaged mean and standard deviation of the feature maps during training, which are completely different between the secret and stego learning tasks. Therefore, when using BN, we store the corresponding feature statistics of the secret DNN model as the side information to be hidden into the stego DNN model with $\mathbf{B}$. Other types of normalization layers, \textit{i.e.} GN, LN, IN, do not introduce additional overhead.

\textbf{Fully Connected Layer}.~
Fully connected layers are very common in DNN models, We treat them as special convolutional layers. For a fully connected layer with $f$ neurons, we represent it as $f$ filters with each being a $c\times1\times1$ tensor, where $c$ is the number of backward neurons that are fully connected to the layer.

\textbf{Output Layer Adaptation}.~In real world applications, the dimension of the output layers are usually different between the secret DNN model and the stego DNN model. For simplicity, we denote the output layers of the secret and stego DNN model as $layer_e$ and $layer_t$, respectively. We further denote the number of neurons in $layer_e$ and $layer_t$ as $O_e$ and $O_t$. When $O_e<O_t$, we upsample $layer_e$ to make it the same dimension as that of $layer_t$ for adaptation. When $O_e>O_t$, however, it is not appropriate to directly downsample $layer_e$ because such a strategy will inevitably distort the function of the secret DNN model. For remedy, we propose to add additional neurons into $layer_e$ to disguise it as the second to last fully connected layer. Then, we newly add a final layer for adaptation, which has the same dimension as that of $layer_t$. Fig. \ref{fig:op} illustrates our output layer adaptation for different cases. All the newly added parameters will not be selected into $\Theta^\mathrm{S}$ to perform the secret task.

\section{Experiments}
\subsection{Setup}
We conduct two types of experiments for our proposed method: 1) steganography of steganographic networks (SSN), where the secret DNN model is a secret decoder, and 2) steganography of general networks (SGN), where the secret DNN model is a DNN model for a general machine learning task. 

In SSN, we use the secret decoder of HiDDeN \cite{zhu2018hidden} as the secret DNN model for the secret task, and we randomly select 11000 images from the COCO dataset  \footnote{https://cocodataset.org/\#home} to form the secret dataset which is split into 10000/1000 images for training and testing. We use the bit error rate (BER) between the original and decoded secret information as the performance indicator of the secret decoder. We take the GTSRB \cite{stallkamp2012man} classification task as the stego task for model disguising. We use the GTSRB dataset as the stego dataset, where 80\%/20\% of the images are randomly selected for training/testing. We adopt the classification accuracy (ACC) to be the indicator to evaluate the performance of the stego DNN model.

In SGN, we evaluate our method on two well-known DNN models: ResNet18 \cite{he2016deep} and U-net \cite{ronneberger2015u}. For ResNet18, we assume the Fashion-MNIST \cite{xiao2017fashion} classification task as the secret task and take the CIFAR10 \cite{krizhevsky2009learning} classification task as the stego task for model disguising. For both the secret and stego datasets, we use their default partitions for training and testing with ACC being the performance indicator. 

For U-net, we assume the image denoising task as the secret task, where the U-net is trained/tested on a secret dataset with 10000/1000 images randomly selected from ImageNet \cite{deng2009imagenet} after adding Gaussian noise. We compute the peak single noise ratio (PSNR) to evaluate the denoising performance. We take image segmentation task as the stego task for model disguising, where the oxford-pet dataset \cite{parkhi2012cats} is adopted as the stego dataset. We randomly separate the Oxford-Pet dataset into three parts, including 6000 images for training, 1282 images for validating and 100 images for testing. We employ the Mean Intersection over Union (mIOU) as the performance indicator for the stego DNN model.

For both SSN and SGN, we set $\lambda_g$=0.01, $\lambda_e$=$\lambda_t$=0.001, $\lambda_p$=0.9, $\tau_{st}$=0.01. For $\tau_{se}$, we set 0.0001, 0.01, and 0.5 for the secret decoder of HiDDeN, ResNet18 and U-net, respectively. We reinitialize the remaining filters by Kaiming initialization \cite{he2015delving}. All the models take the BN as normalization layer and are optimized using Adam \cite{kingma2014adam} optimizer. All our experiments are conducted on Ubuntu 18.04 system with four NVIDIA
RTX 1080 Ti GPUs.

\begin{table}[t]
\renewcommand{\arraystretch}{1.0} 
  \centering

  \resizebox{\linewidth}{!}{
\begin{tabular}{c|c|c|c}
    \hline
    \multirow{2}{*}{Secret} & SSN                        & \multicolumn{2}{c}{SGN} \\
    \cline{2-4}
    \multirow{2}{*}{DNN}    &   HiDDeN-                   & Fashion-MNIST-                    &  ImageNet-    \\
    \multirow{2}{*}{Model} &   Secret Decoder                 & Classification            &  Denoising   \\
                          &   BER \ $(\times10^{-5})$                   &  ACC(\%)                  &   PSNR(dB)            \\ 
    \hline
    Original          &   6.68                      & 93.99                     &   28.98               \\
    \hline
    Recovered    &   6.68                      & 93.80                     &   28.92               \\
\hline

\end{tabular}}
\caption{The recoverability of the proposed method.}
  \label{tab:Recoverability}
\end{table}

\begin{table}[t]
\renewcommand{\arraystretch}{1.0} 
  \centering

  \resizebox{0.91\linewidth}{!}{
\begin{tabular}{c|c|c|c}
    \hline
                             & SSN                        & \multicolumn{2}{c}{SGN} \\
    \cline{2-4}
    \multirow{1}{*}{DNN} &   GTSRB-                  & CIFAR10-                 &  Oxford-Pet-    \\
    \multirow{1}{*}{Model} &   Classification            & Classification            &  Segmentation   \\
                          &   ACC (\%)                  &  ACC (\%)                 &  mIOU (\%)            \\ 
    \hline
    Cover \quad    &   99.97                      & 93.46                     &   87.68               \\
    \hline
    Stego  \quad     &   99.81                   & 93.02                & 87.05          \\
    \hline
\end{tabular}}
\caption{The fidelity of the proposed method.}
  \label{tab:Fidelity}
\end{table}

\begin{table}[t]
\renewcommand{\arraystretch}{1.0} 
  \centering

  \scalebox{0.84}{
  \resizebox{\linewidth}{!}{
\begin{tabular}{c|c|c|c}     
    \hline
   \multirow{2}{*}{Expansion}             &   SSN                       & \multicolumn{2}{c}{SGN}    \\
    \cline{2-4}
    
    \multirow{2}{*}{Rate}  &   HiDDeN-            & \multirow{2}{*}{ResNet18}              & \multirow{2}{*}{U-Net}     \\
                     &   Secret Decoder            &                       &     \\
    \hline
    $e \ (\times10^{-3})$ &   1.65                 & 0.00        & 0.14               \\
\hline
\end{tabular}}}
\caption{The capacity of the proposed method.}
    
  \label{tab:Capacity}
\end{table}

\subsection{Recoverability} We evaluate the recoverability of our proposed method according to the performance of the secret DNN model before and after the model disguising, i.e., the performance of the original and recovered secret DNN models (from the stego DNN model). Table \ref{tab:Recoverability} gives the results for different steganographic scenarios. It can be seen that, the performance of the secret decoder is not affected at all after model disguising in the SSN case. For SGN, the performance of the secret DNN models slightly decrease, which does not affect the function of the DNN models for the secret tasks.

\subsection{Fidelity}
To evaluate the fidelity of the stego DNN model, we train a cover DNN model which has the same architecture as the stego DNN model for the stego task. The training is conducted on the training set of the stego dataset which has been used for model disguising, where all the parameters in the cover DNN model are tuned to achieve the best performance. Furthermore, we use the same test set to evaluate the performance of the two models.

Table \ref{tab:Fidelity} presents the performance of the stego DNN model and the cover DNN model for both SSN and SGN. For SSN, it can be seen that the ACC of the stego DNN model is reduced with less than $0.2\%$ after the model disguising. The results of the SGN are similar to those of SSN. In particular, after model disguising, the ACC/mIOU of the stego DNN model is slightly degraded when compared with the corresponding cover DNN model on the CIFAR100/Oxford-Pet dataset. These results indicate that the fidelity of our stego DNN model is favorable, which has little performance degradation caused by model disguising on the stego task.

\subsection{Capacity}
The capacity of the proposed method is evaluated in terms of the expansion rate defined as
\begin{equation}
e=\frac{N_{stego} }{N_{sec}}-1, 
\end{equation}
where $N_{sec}$ and $N_{stego}$ refer to the number of parameters in the original secret DNN model and the corresponding stego DNN model, respectively. Thus, we have $e \geq 0$, which is closely related to the number of parameters added when adjusting the output layer.  The lower the expansion rate, the higher the capacity, meaning that there is less overhead created for transmission after the model disguising. Table \ref{tab:Capacity} gives the expansion rates $e$ for different steganographic scenarios. It can be seen that, for both SSN and SGN, the expansion rates are extremely small with less that $2 \times 10^{-3}$. In the ResNet18 case, the output layers are with the same dimension for the secret and stego tasks, so the parameters do not expand at all with $e=0$.

\begin{table}[t]
\renewcommand{\arraystretch}{1.0} 
  \centering    

  \scalebox{0.61}{
  \resizebox{\linewidth}{!}{
\begin{tabular}{c|c}
    \hline
    Classifier & Detection Accuracy (\%) \\
    \hline
     SVM   & 47.50 \\
    \hline
     MLP   & 50.00 \\
\hline
\end{tabular}}}
\caption{The undetectability of the proposed method.}
  \label{tab:Undetectability}
\end{table}

\begin{table*}[t]
\renewcommand{\arraystretch}{1.0} 
  \centering

  \resizebox{\linewidth}{!}{
\begin{tabular}{c|c|c|c|c}
\hline
           \multirow{2}{*}{Category}     & \multirow{2}{*}{Method}  &\multirow{2}{*}{Capacity}   &  Recoverability &  Undetectability \\
                &        &   & BER(\%) / PSNR(dB)   &   $P_\mathrm{E}$(\%) / Detection accuracy (\%)\\ 
            
           \hline
                              & Tang \textit{et al}.\cite{tang2017automatic}&200.00MB (0.40bpp)    &0 / - &17.40 / -    \\
                              &Yang \textit{et al}. \cite{yang2019embedding}&160.00MB (0.50bpp)    &0 / - &29.71 / -    \\
\multirow{1}{*}{Traditional}   &Liao \textit{et al.} \cite{liao2019new} & 160.00MB (0.50bpp)    &0 / -    & 25.83 / -\\
\multirow{1}{*}{steganography}  &Tang \textit{et al}. \cite{tang2020automatic}& 200.00MB (0.40bpp)   &0 / - &32.19 / -    \\
                                  &Lu \textit{et al.} \cite{8789545}       & 655.35MB (0.015bpp)    &0 / -    & 34.17 / -\\
                                 &Liao \textit{et al.} \cite{9124671}       & 160.00MB (0.50bpp)    &0 / -     & 35.40 / -\\

\hline
                            &Zhu \textit{et al}. \cite{zhu2018hidden}     & 10.00MB (8.00bpp)  &- / 35.70  & - / 76.49\\
\multirow{1}{*}{DNN-based} &Bakuja \textit{et al}. \cite{baluja2019hiding}& 10.00MB (8.00bpp) &- / 34.13 &- / 99.67    \\
\multirow{1}{*}{steganography}   &Weng  \textit{et al}. \cite{weng2019high}      &  10.00MB (8.00bpp) &- / 36.48 &- / 75.03    \\ 
                                &Jing  \textit{et al}. \cite{jing2021hinet}     & 10.00MB (8.00bpp)  &- / 46.78 &- / 55.86     \\
\hline
                                &\multirow{3}{*}{\textbf{Ours}}       & \multirow{3}{*}{10.00MB ($e$=0.00)}     &   Performance reduction   &  \multirow{2}{*}{- / 47.50 (SVM)} \\ 
                                &           &  & 0.00\% in BER, 0.19\% in  &\multirow{2}{*}{- / 50.00 (MLP)} \\
                                &           &  & ACC, 0.06dB in PSNR &  \\
\hline
\end{tabular}}
\caption{Performance comparisons among different schemes for hiding DNN models.}
  \label{tab:comparison}
\end{table*}
\subsection{Undetectability}
\label{un}
To evaluate the undetectability, we first establish a model pool with different stego DNN models and cover DNN models trained. We select five benchmark datasets: MNIST \cite{lecun1998gradient}, Fashion-MNIST, CIFAR10, CIFAR100, GTSRB to generate $\mathrm{A}^2_5 = 20$ different secret-stego dataset pairs. Then, we train five well-known DNN models: LeNet-5 \cite{lecun1998gradient}, AlexNet \cite{krizhevsky2012imagenet}, VGG11, VGG19 \cite{simonyan2015vgg} and ResNet34 on the 20 dataset pairs to generate 100 different stego DNN models. Since we can only obtain 25 cover DNN models on the five datasets and five DNN models, we train each cover DNN model four times with different initialized parameters for augmentation. Take the three pairs of cover and stego DNN models (HiDDeN-Secret Decoder/ResNet18/U-net) previously trained into consideration, we obtain a total of 103 stego DNN models and 103 cover DNN models in our model pool.

Next, we compute a 100 bin-histogram from the parameters in each model to get a 100-dimensional feature vector for training a support vector machine (SVM) and a multilayer perceptron (MLP) classifier. Specifically, we randomly select 83 pairs of such features from the cover and stego DNN models for training and the remaining 20 pairs are used for testing. The detection accuracy are shown in Table \ref{tab:Undetectability}, we can see that both the SVM and MLP classifiers are not able to detect the existence of the secret DNN models from the stego DNN models with detection accuracy close to $50\%$. 
 
\subsection{Comparisons}

In this section, we compare our proposed method with the existing state-of-the-art steganographic schemes for hiding DNN models in terms of capacity, recoverability, and undetectability. For fair comparisons among different types of steganographic schemes, we assume the data to be hidden is a secret DNN model with size of 10MB for all the methods. The capacity is evaluated as the size of stego data, the recoverability is reported as the BER or PSNR of the recovered secrets for the existing schemes, the undetectability is indicated as the average detection error $P_{\mathrm{E}}$ (which is the mean of false alarm rate and missed detection rate) or the detection accuracy (50\% means random guess).

Table \ref{tab:comparison} reports the comparisons among different schemes. For the traditional steganographic schemes \cite{liao2019new, 8789545, 9124671, tang2017automatic,yang2019embedding, tang2020automatic}, we duplicate the $P_{\mathrm{E}}$ reported at a certain payload for undetectability from each paper, where the payload is the bit per pixel (bpp), i.e., how many bits can be embedded in an 8-bit gray-scale pixel. For the DNN-based steganographic schemes \cite{baluja2019hiding, zhu2018hidden, weng2019high, jing2021hinet}, we duplicate the PSNR (on imagenet) and detection accuracy at a payload of 8bpp from the results reported in \cite{jing2021hinet}. Here, we assume the 10MB secret data is a natural image with meaningful content for these DNN steganographic schemes.

It can be seen from Table \ref{tab:comparison} that the size of our stego data is less than 1/10 of size of the stego data that are required using the existing traditional steganographic schemes. Compared with the existing DNN-based steganographic schemes \cite{zhu2018hidden, baluja2019hiding, weng2019high, jing2021hinet}, which are not able to losslessly recover the secret and are not suitable for hiding DNN models, our scheme achieves better undetectability (closer to 50\%). In addition, all the existing DNN-based steganographic schemes involve the change of data format between the secret-data (a DNN model) and the stego data (images).

To demonstrate that the existing DNN-based steganographic schemes are not suitable for hiding DNN models, we further adopt a recent DNN-based steganographic scheme
 \cite{jing2021hinet} to hide two DNN models: the HiDDeN secret decoder and the ResNet18.
Specifically, we convert the parameters of a secret DNN model to a set of binary streams to form a set of meaningless images with the size of $256 \times 256 \times 3$. Next, we embed each image into a cover image randomly selected from the COCO dataset using the encoder proposed in \cite{jing2021hinet}, which produces a stego image for decoding. Finally, we recover the images from the stego images using the corresponding decoder, from which we restore the DNN model. Table \ref{tab:NtoI} gives the performance of the recovered DNN models using our proposed method and the method proposed in \cite{jing2021hinet}. It can be seen that, by using the DNN-based steganographic scheme, the recovered DNN models do not work at all.

\begin{table}[t]
\renewcommand{\arraystretch}{1.0} 
  \centering

  \scalebox{0.9}{
  \resizebox{\linewidth}{!}{
\begin{tabular}{c|c|c}
    \hline 
     \multirow{2}{*}{Steganographic} & HiDDeN- &  \multirow{2}{*}{ResNet18}                        \\
    \multirow{2}{*}{Method} & Secret Decoder &        \\
                             & BER (\%) &   ACC(\%)  \\
    \hline
    Jing  \textit{et al}. \cite{jing2021hinet}  &  50.13 &  10.00  \\
    \hline
     \textbf{Ours}      &        6.68 $\times 10^{-3}$   &  93.80   \\

\hline
\end{tabular}}}
\caption{The performance of the recovered DNN models using different schemes.}
  \label{tab:NtoI}
\end{table}




\section{Conclusion}

In this paper, we propose a novel steganographic scheme to tackle the problem of covert communication of secret DNN models. Our scheme is able to disguise a secret DNN model into a stego DNN model which performs an ordinary machine learning task without being noticed. And the secret DNN model can be recovered from the stego DNN model on the receiver side. In model disguising, we first select a set of filters from the secret DNN model, which are important on the secret task but trivial on the stego task. We fine tune these filters on the secret task and then reactivate the other filters for the stego task according to a partial optimization strategy to train a stego DNN model. These are progressively done to obtain a stego DNN model that works well on both the secret and stego tasks. Various experiments have been conducted to demonstrate the advantage of our proposed method for covert communication of the DNN models, which achieves satisfactory performance in terms of recoverability, fidelity, capacity and undetectability.

\section{Acknowledgments}
This work was supported in part by the National Natural Science Foundation of China under Grants 62072114, U20A20178, U20B2051, U1936214, in part by the Project of Shanghai Science and Technology Commission 21010500200.

\bibliography{aaai23}



\end{document}